\documentclass[superscriptaddress,showpacs,aps,twocolumn]{revtex4}
\usepackage{amssymb}
\usepackage{amsmath}
\usepackage{graphicx}
\usepackage{color}
\usepackage{amsfonts}
\usepackage{multirow}

\begin{document}

\title{Van der Waals effects on grazing incidence fast atom diffraction for
H/LiF(001)}
\author{G. A. Bocan\thanks{%
Author to whom correspondence should be addressed. \newline
Electronic address: gisela.bocan@cab.cnea.gov.ar}}
\affiliation{Consejo Nacional de Investigaciones Cient\'ificas y T\'ecnicas (CONICET) and
Centro At\'omico Bariloche, Av. Bustillo 9500, 8400 S.C. de Bariloche,
Argentina.}
\author{J. D. Fuhr}
\affiliation{Consejo Nacional de Investigaciones Cient\'ificas y T\'ecnicas (CONICET) and
Centro At\'omico Bariloche, Av. Bustillo 9500, 8400 S.C. de Bariloche,
Argentina.}
\author{M. S. Gravielle}
\affiliation{Instituto de Astronom\'{\i}a y F\'{\i}sica del Espacio (IAFE, CONICET-UBA),
Casilla de Correo 67, Sucursal 28, C1428EGA Buenos Aires, Argentina.}

\begin{abstract}
We theoretically address grazing incidence fast atom diffraction (GIFAD) for
H atoms impinging on a LiF(001) surface. Our model combines a description of
the H-LiF(001) interaction obtained from Density Functional Theory
calculations with a semi-quantum treatment of the dynamics. We analyze
simulated diffraction patterns in terms of the incidence channel, the impact
energy associated with the motion normal to the surface, and the relevance
of Van der Waals (VdW) interactions. We then contrast our simulations with
experimental patterns for different incidence conditions. Our most important
finding is that, for normal energies lower than $0.5$ eV and incidence along
the $\langle 100\rangle $ channel, the inclusion of Van der Waals
interactions in our potential energy surface yields a greatly improved
accord between simulations and experiments. This agreement strongly suggests
a non-negligible role of Van der Waals interactions in H/LiF(001) GIFAD in
the low-to-intermediate normal energy regime.
\end{abstract}

\maketitle
\date{\today}

%%%%%%%%%%%%%%%%%%%%%%%%%%%%%%%%%%%%%%%%%%%%%%%%%%%%%%%%%%%%%%%%%%%%%

\section{Introduction}

The extraordinary sensitivity of grazing incidence fast atom diffraction
(GIFAD or FAD) has turned this phenomenon into a powerful surface analysis
technique, positioning it also as a most useful tool for testing potential
energy surfaces (PESs) ~\cite%
{Schuller2009d,Schuller2010,Rubiano2013,Zugarramurdi2015}.

When atomic projectiles in the keV energy range grazingly impinge on a
crystal surface along a low-index crystallographic direction, scattering
proceeds under axial surface channeling conditions \cite{Winter2011}. The fast
motion along the channel is, on a first approach, sensitive only to the
periodic-PES average in this direction. The associated energy $E_{\parallel
} $ is thus essentially conserved, and motions parallel and perpendicular to
the channel get decoupled from each other. The scattering process can then
be projected into the plane normal to the channeling direction, where motion
proceeds with an energy $E_{\perp }$ in a hyperthermal up to eV energy
regime and a \mbox{De Broglie} wavelength of the order of the interatomic
spacing. Diffraction patterns arise due to the interference produced by the
periodic array of channels, modulated by that originated within a given
channel \cite{Schuller2008,Schuller2009}. GIFAD was first reported for light
projectiles impinging on wide band-gap insulating surfaces ~\cite%
{Schuller2007,Rousseau2007}, but has since been observed for a variety of
systems including semiconductors ~\cite{Khemliche2009,Debiossac2014}, metals
~\cite{Bundaleski2008,Busch2009}, adsorbate-covered metal surfaces ~\cite%
{Schuller2009b}, ultrathin films ~\cite{Seifert2010} and organic molecules
on metal substrates ~\cite{Seifert2013}.

From the theoretical standpoint, the extreme sensitivity of GIFAD poses a
challenge for achieving an appropriate description of this phenomenon. The
construction of a projectile-surface potential which includes the key
features of the interaction and a scattering dynamics representation which
retains the quantum character of the process are necessary ingredients for
attaining good accord with experimental diffraction patterns.

Projectile-surface interaction potentials for GIFAD simulations are usually
built from Density Functional Theory (DFT) calculations. Within a standard
DFT approach, the exchange-correlation energy is described by means of local
or semi-local functionals, thus leaving long-range dispersion forces aside.
This level of approximation is appropriate for high-enough $E_{\perp }$
GIFAD, but it may not be sufficient for the low-$E_{\perp }$ regime. In this
latter case, the projectile is scattered farther from the surface, in lower
electron-density regions where Van der Waals (VdW) interactions should not
be neglected.

Studies on the dynamic aspects of VdW interactions are still scarce ~\cite%
{Debiossac2014,Boereboom2013,Wijzenbroek2014,MGondre2015,Schuller2012}. In
GIFAD literature, Zugarramurdi \textit{et al. }~\cite{Zugarramurdi2015}
modeled the interaction of He atoms with a graphene layer on 6H-SiC(0001) by
means of a pairwise additive Lennard-Jones potential fitted from Helium Atom
Scattering (HAS) data; Debiossac \textit{et al.} ~\cite{Debiossac2014}
considered ad-hoc corrections to the attractive part of a DFT potential for
He atoms impinging on the $\beta _{2}(2\times 4)$ reconstructed GaAs(001)
surface; and Sch\"{u}ller \textit{et al.} ~\cite{Schuller2012} included VdW
contributions into a DFT interaction potential by means of Grimme's
semi-empirical approach ~\cite{Grimme2006}. The latter authors addressed the
role of VdW contributions for He atoms impinging on the insulating MgO(001)
surface, reporting no significant effect due to VdW either on the rumpling,
the interaction potential $V(z)$, or the corrugation of the equipotential
curves in the  normal energy range relevant for GIFAD.

In this article, we report on VdW effects on GIFAD for H/LiF(001), one of
the systems where this phenomenon was initially observed ~\cite%
{Schuller2007,Rousseau2007}. Despite the considerable amount of experimental
data for this collision system ~\cite%
{Schuller2007,Rousseau2008,Lienemann2011,Winter2011,Busch2012,Winter2014},
theoretical research is scarce ~\cite{Muzas2015,Muzas2016}, and ab initio
simulations that satisfactorily reproduce all the experiments are still
lacking.

We describe the elastic scattering of H atoms off the LiF(001) surface
within the Surface-Initial Value Representation (SIVR) approximation ~\cite%
{Gravielle2014}, which is a semi-quantum method that affords a clear
representation of the main physical mechanisms in terms of classical
trajectories through the Feynman path integral formulation of quantum
mechanics ~\cite{Miller2001}. The SIVR method includes an approximate
representation of classically forbidden transitions on the dark side of the
rainbow angle, providing an appropriate description of GIFAD patterns along
the whole angular range ~\cite{Gravielle2014,Gravielle2015}. Another
noteworthy point is that, with a relatively low computational cost, the SIVR
approach takes into account the three-dimensionality of the PES, without
averaging it along the incidence direction. These features make SIVR a most
attractive alternative to quantum wave packet propagation methods.

We show that a description of the H-Surface interaction based on PAW
pseudopotentials and the semi-local GGA-PBE functional fails in reproducing
the diffraction patterns for incidence along the $\langle 100\rangle $
channel in the \mbox{$E_{\perp}\,\lesssim\,0.5\,{\rm eV}$} energy range.
However, upon inclusion of VdW interactions in the PES through the
semi-empirical approach by Grimme ~\cite{Grimme2006}, we achieved a much
improved, almost quantitative accord between our simulations and the
experiments. To our knowledge, this agreement provides the first indication
of the non-negligible role of VdW interactions in GIFAD, obtained from a non
ad-hoc potential.

The paper is organized as follows: The theoretical models used to describe
the quantum scattering and the projectile-surface interaction are summarized
in Sec. II. Results for incidence along the $\langle 110\rangle$ and $%
\langle 100\rangle$ channels are presented and discussed in Sec. III, with
the focus on the influence of the VdW contribution. In Sec. IV we outline
our conclusions.

%%%%%%%%%%%%%%%%%%%%%%%%%%%%%%%%%%%%%%%%%%%%%%%%%%%%%%%%%%%%%%%%%%%%%%%%%%%%%%%

\section{Theoretical Model}

Our theoretical description of GIFAD combines a semi-quantum representation
of the scattering process with an accurate projectile-surface interaction
potential. They are both summarized in the following subsections.

%%%%%%%%%%%%%%%%%%%%%%%%%%%%%%%%%%%%%%%%%%%%%%%%%%%%%%%%%%%%%%%%%%%%%%%%%%%%%%%

\subsection{Scattering process}

We treat the scattering dynamics of H atoms grazingly colliding with the
LiF(001) surface by means of the SIVR approximation ~\cite%
{Gravielle2014,Gravielle2015}, expressing all quantities in atomic units
(a.u.). Within this approach the transition amplitude per unit of surface
area $\mathcal{S}$ reads ~\cite{Gravielle2014}
\begin{equation}
A_{if}^{(SIVR)}=\ \frac{m_{P}K_{i}}{\mathcal{S}}\int\limits_{\mathcal{S}}d%
\overrightarrow{R}_{os}\ \int d\Omega _{o}\ a_{if}^{(SIVR)}(\overrightarrow{R%
}_{o},\overrightarrow{K}_{o}),  \label{Aif-sivr}
\end{equation}%
where $\overrightarrow{K}_{i}$ is the initial momentum of the impinging
atom, $K_{i}=\left\vert \overrightarrow{K}_{i}\right\vert $, $m_{P}$ is the
projectile mass, and $a_{if}^{(SIVR)}(\overrightarrow{R}_{o},\overrightarrow{%
K}_{o})$ is the partial transition amplitude associated with the classical
projectile path $\overrightarrow{\mathcal{R}}_{t}\equiv \overrightarrow{%
\mathcal{R}}_{t}(\overrightarrow{R}_{o},\overrightarrow{K}_{o})$, with $%
\overrightarrow{R}_{o}$ and $\overrightarrow{K}_{o}$ respectively being the
starting ($t=0$) position and momentum of the projectile. In Eq. (\ref%
{Aif-sivr}), the starting position is expressed as $\overrightarrow{R}_{o}=%
\overrightarrow{R}_{os}+Z_{o}\widehat{z}$, where $\overrightarrow{R}%
_{os}=X_{o}\widehat{x}+Y_{o}\widehat{y}$ is the component parallel to the
surface plane, $\widehat{z}$ is the normal to the surface and  $Z_{o}$ is a
reference distance for which the projectile is hardly affected by the
surface interaction. In turn, the starting momentum $\overrightarrow{K}_{o}$%
, with $\left\vert \overrightarrow{K}_{o}\right\vert =K_{i}$, aims in the
direction of the solid angle $\Omega _{o}$, which varies around $\widehat{K}%
_{i}=\overrightarrow{K}_{i}/K_{i}$.

The partial transition amplitude $a_{if}^{(SIVR)}$ can be expressed as
\begin{eqnarray}  \label{aif}
a_{if}^{(SIVR)}(\overrightarrow{R}_o,\overrightarrow{K}_o) &=&\
-\int\limits_0^{+\infty }\frac{\left\vert J_M(t)\right\vert
^{1/2}e^{i\nu_t\pi /2}}{(2\pi i)^{9/2}}V_{PS}(\overrightarrow{\mathcal{R}}_t)
\notag \\
&&\times \exp \left[ i\left( \varphi_t^{(SIVR)}-\overrightarrow{Q} \cdot
\overrightarrow{R}_o\right) \right]\,dt,  \notag \\
\end{eqnarray}%
where $V_{PS}$ represents the projectile-surface interaction, $%
\overrightarrow{Q}=\overrightarrow{K}_f-\overrightarrow{K}_i$ is the
projectile momentum transfer, with $\overrightarrow{K}_f$ the final
projectile momentum satisfying energy conservation, i.e. $K_f=K_i$, and
\begin{equation}
\varphi_t^{(SIVR)}=\int\limits_0^{t}\left[ \frac{1}{2m_P}\left(
\overrightarrow{K}_f- \overrightarrow{\mathcal{P}}_{t^{\prime}}\right)^{2}-
V_{PS}(\overrightarrow{\mathcal{R}}_{t^{\prime }})\right]\, dt^{\prime }
\label{fitot}
\end{equation}%
is the SIVR phase at the time $t$, with $\overrightarrow{\mathcal{P}}%
_{t}=m_{P}\,d\overrightarrow{\mathcal{R}}_{t}/dt$ the classical projectile
momentum. In Eq. (\ref{aif}) the Maslov function ~\cite{Guantes2004}
\begin{equation}
J_M(t)=\det \left[ \frac{\partial \overrightarrow{\mathcal{R}}_t (%
\overrightarrow{R}_o,\overrightarrow{K}_o)}{\partial \overrightarrow{K}_o}%
\right] =\left\vert J_M(t)\right\vert e^{i\nu_t\pi}  \label{J}
\end{equation}%
is a Jacobian factor (a determinant) evaluated along the classical
trajectory $\overrightarrow{\mathcal{R}}_t$, with $\left\vert
J_M(t)\right\vert$ the modulus of $J_M(t)$ and $\nu_t$ an integer number
that accounts for the sign of $J_M(t)$ at a given time $t$, increasing by $1$
every time that $J_M(t)$ changes its sign along the trajectory.

The SIVR differential probability for elastic scattering with final momentum
$\overrightarrow{K}_{f}$ in the direction of the solid angle $\Omega
_{f}\equiv (\theta _{f},\varphi _{f})$ is derived as $dP^{(SIVR)}/d\Omega
_{f}=K_{f}^{2}\left\vert A_{if}^{(SIVR)}\right\vert ^{2}$ ~\cite%
{Gravielle2014}, with $\theta _{f}$ the final polar angle, measured with
respect to the surface, and $\varphi _{f}$ the azimuthal angle, measured
with respect to the channel direction (see Fig. \ref{Geom1}). In the present
work, the transition amplitude $A_{if}^{(SIVR)}$ is obtained from Eq. (\ref%
{Aif-sivr}) by employing the MonteCarlo technique with more than $4\times
10^{5}$ points in the $\overrightarrow{R}_{os}$ and $\Omega _{o}$ integrals.
In such integrations, the random $\overrightarrow{R}_{os}$ values are
derived from a Gaussian distribution covering an area $\mathcal{S}$ equal to
$2$ or $3$ reduced unit cells, while the $\Omega _{o}$ values are obtained
from a Gaussian distribution encompassing an angular region determined by
the Heisenberg uncertainty relation. In this aspect, it should be mentioned
that the $\overrightarrow{R}_{os}$ and $\Omega _{o}$ distributions are in
principle defined by the profile of a coherent wave packet associated with
the impinging particle, which depends on the collimation of the incident
beam ~\cite{Seifert2015,Gravielle2015,Gravielle2016}. Here we have
considered standard sizes of the $\overrightarrow{R}_{os}$ and $\Omega _{o}$
distributions because reported experimental data ~\cite%
{Schuller2007,Rousseau2008,Lienemann2011,Winter2011,Busch2012,Winter2014}
lack information about collimating parameters. In addition, the starting
normal distance was chosen as $Z_{o}=1.4\ a$ ($a$ is the lattice constant),
to ensure a negligible projectile-surface interaction.

%%%%%%%%%%%%%%%%%%%%%%%%%%%%%%%%%%%%%%%%%%%%%%%%%%%%%%%%%%%%%%%%%%%%%%%%%%%%%%%

\subsection{Projectile-surface potential}

To obtain the H-LiF(001) potential $V_{PS}$ required in Eqs. (\ref{aif}) and
(\ref{fitot}), we make use of DFT, as implemented in the QUANTUM ESPRESSO
code ~\cite{Giannozzi2009}, to calculate the system's energy for a grid of $%
(X_{i},Y_{i},Z_{i})$ positions of the H atom over a relaxed LiF(001)
surface. The grid is three-dimensional (3D) and is built out of a selection
of 6 high-symmetry $(X_{i},Y_{i})$ configurations and 62 $Z_{i}$ values ($Z=0
$ falls on the topmost F layer). We then apply an interpolation technique,
which combines the Corrugation Reducing Procedure (CRP) ~\cite{Busnengo2000}
with the cubic spline method, to obtain the potential energy for an
arbitrary $(X,Y,Z)$ position.

For the DFT calculations, we use projector augmented-wave (PAW)
pseudopotentials ~\cite{Kreese1999,PseudoPots} to describe the electron-core
interaction, while for the exchange-correlation functional we consider two
different models: (a) the generalized gradient approximation (GGA), with the
Perdew-Burke-Ernzerhof (PBE) functional ~\cite{Perdew1996} (henceforth
referred to as PAW-PBE), or (b) the DFT-D2 approach of Grimme ~\cite%
{Grimme2006,Barone2009}, which introduces a semi-empirical correction to the
GGA functional to account for long-range VdW interactions (henceforth
PAW-PBE-VdW).

For both PAW-PBE and PAW-PBE-VdW models, we choose relevant parameters of
the DFT calculations so that ab-initio energies differ from the converged
result in less than 5 meV. The energy cutoff in the plane-wave expansion is
100 Ryd for the wave functions and 1000 Ryd for the charge density and
potential; fractional occupancies are determined through a gaussian
broadening approach with $\sigma\,=\,10^{-6}$ Ryd.; and the Brioullin-zone
integration is performed with a $4\times4\times1$ Monkhorst-Pack grid of
special k-points with an offset. The LiF lattice constant is $a\,=\,4.066$
\AA {} for the PAW-PBE case and $a\,=\,4.063$ \AA {} for the PAW-PBE-VdW
case, both slightly higher than the experimental value of 4.02 \AA {} ~\cite%
{Ekinci2004}.

We represent the LiF(001) surface by means of the supercell-slab scheme. The
supercell consists of a $2\times 2$ surface cell, a five-layer slab and a
vacuum distance of $5\,a$. The surface equilibrium geometry is reached by
relaxing the two topmost LiF(001) planes from their bulk positions. F and Li
atoms initially in the same plane relax differently and the resulting
surface geometry presents a \textit{rumpling}, which we define as the
distance between relaxed F and Li planes. For the topmost F and Li planes,
we get rumplings of $+0.070$ \AA {} (PAW-PBE) and $+0.088$ \AA {}
(PAW-PBE-VdW), with F atoms moving outward and Li atoms moving inward. These
values are consistent with LEED experiments which yield a rumpling of $%
0.036\,\pm \,0.1$ \AA {} ~\cite{Vogt2002} and, particularly, our PAW-PBE
result compares very well with Vogt's $0.068$ \AA {} ~\cite{Vogt2008}, also
obtained from a GGA calculation. The relaxed surface is thereafter kept
frozen both for the energy grid calculations, and during the scattering
process.

%%%%%%%%%%%%%%%%%%%%%%%%%%%%%%%%%%%%%%%%%%%%%%%%%%%%%%%%%%%%%%%%%%%%%%%%%%%%%%%

\section{Results and Discussion}

In this section we first discuss the general features of the PESs used in
this work; then we proceed to a systematic theoretical study of GIFAD
patterns for H/LiF(001) in terms of the incidence channel and the normal
energy $E_{\perp }=E_{tot}\,sin^{2}\theta _{i}$ ($E_{tot}$ the total energy
and $\theta _{i}$ the incidence angle relative to the surface plane. See
Fig. ~\ref{Geom1}); and finally, we compare our theoretical simulations with
experimental GIFAD distributions available in the literature ~\cite%
{Schuller2007,Rousseau2008,Lienemann2011,Winter2011,Busch2012,Winter2014,Muzas2016}%
, in order to probe the PAW-PBE and PAW-PBE-VdW projectile-surface
potentials.

The geometry of GIFAD for H/LiF(001) as well as the channeling directions $%
\langle 110\rangle $ and $\langle 100\rangle $ are illustrated in Fig. \ref%
{Geom1}.

%%%%%%%%%%%%%%%%%%%%%%%%%%%%%%%%%%%%%%%%%%%%%%%%%%%%%%%%%%%%%%%%%%%%%%%%%%%%%%%

\subsection{Analysis of the PES}

The PES profile and corrugation near the reflection region are central in
determining GIFAD patterns. We stress again that our PES is 3D and no
dimension reduction is made during the dynamics. However, the fast motion of
the projectile along the channel is in fact mainly sensitive to the average
of $V_{PS}$ in this direction and thus we will discuss the PES role in these
terms.

\begin{figure}[tbp]
\includegraphics[width=0.5\textwidth]{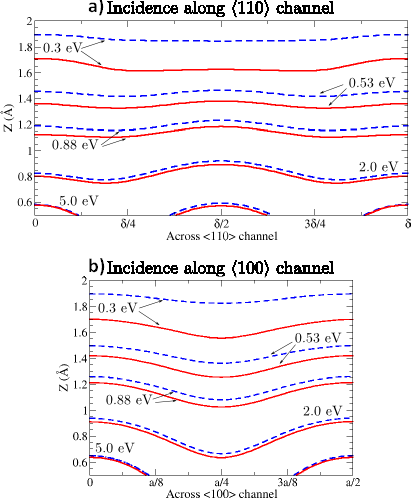}
\caption{(Color online) Equipotential contours averaged along the incidence
channel. Blue dashed curves for the PAW-PBE PES and red solid curves for the
PAW-PBE-VdW PES. a) Incidence along $\langle110\rangle$ direction. b)
Incidence along $\langle100\rangle$ direction. $\protect\delta$ and $a/2$
correspond to the respective channel widths as depicted in Fig. \protect\ref%
{Geom1}}
\label{Equipots}
\end{figure}

For our PAW-PBE and PAW-PBE-VdW PESs, in Figs. \ref{Equipots}a and \ref%
{Equipots}b we consider the energy averages respectively along the $\langle
110\rangle $ and $\langle 100\rangle $ channels, and depict equipotential
contours across them. Across the $\langle 110\rangle $ channel the
equipotential curves have local maxima both at the border and at the middle
of the channel, respectively corresponding to the rows of F and Li ions.
Across the $\langle 100\rangle $ channel, the equipotential curves have only
one maximum at the border of the channel, corresponding to the F-Li rows. In
Table \ref{Tab1}, we consider equipotential curves in the $0.3$-$0.9$ eV
range and show their $Z$-range $\left( Z_{min},Z_{max}\right) $ and
corrugation $\Delta Z\,=\,Z_{max}\,-\,Z_{min}$ for both PESs and channels.
Note that, for projectiles that run parallel to the channel without
suffering any azimuthal deflection (i.e. $\varphi _{f}=0$), these $Z$-ranges
determine the reflection region.

From Fig. \ref{Equipots} and Tab. \ref{Tab1} we obtain that \textit{i)} the
corrugation across the $\langle 100\rangle $ channel is much higher than
that across the $\langle 110\rangle $ channel;\textit{\ ii)} the inclusion
of VdW interactions results in a less repulsive and higher corrugated PES,
these effects growing stronger for lower energies; and \textit{iii)} VdW
interactions seem to play a more important role for the $\langle 100\rangle $
channel than for the $\langle 110\rangle $ one.

\begin{figure}[tbp]
\includegraphics[width=0.5\textwidth]{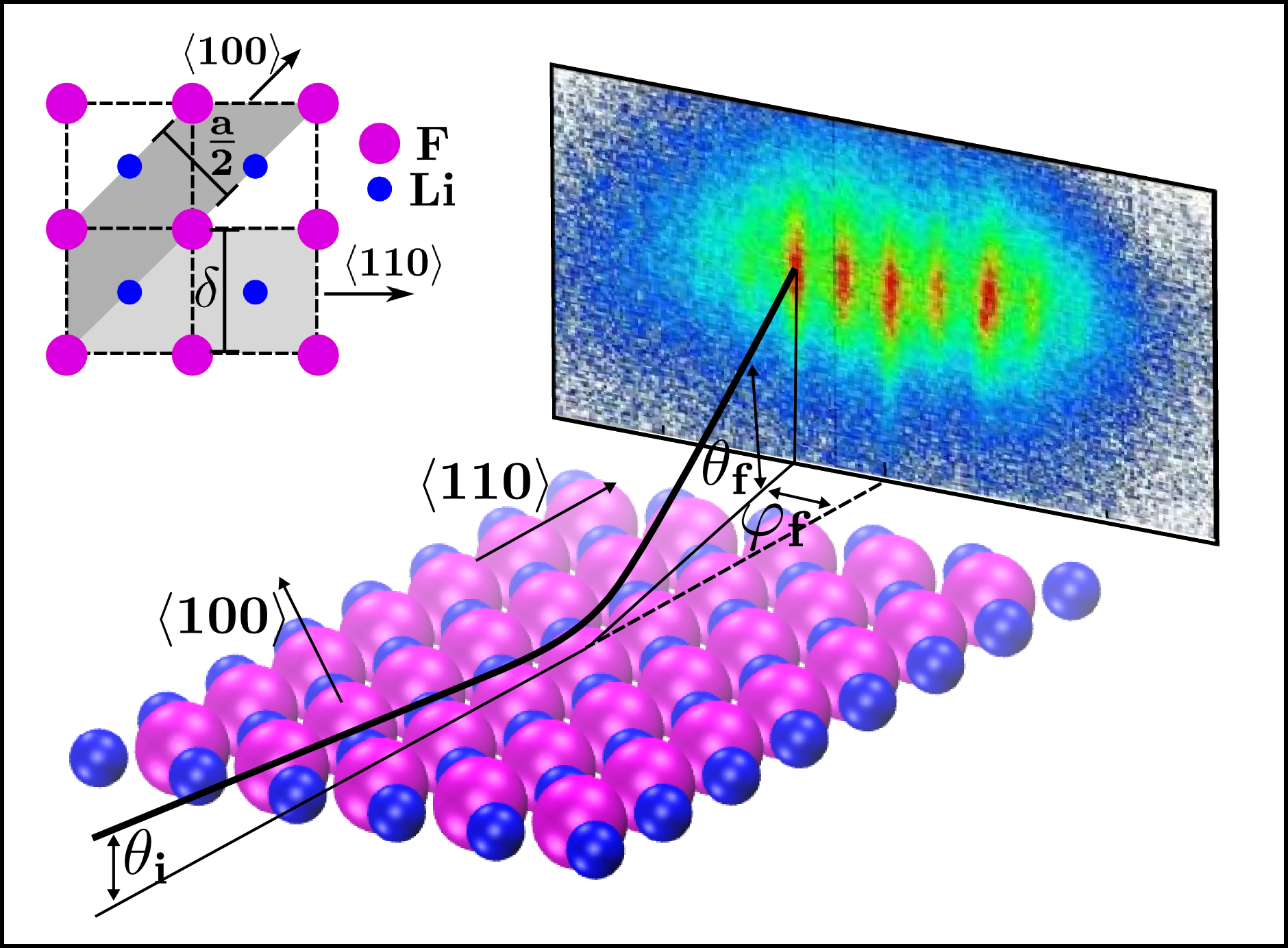}
\caption{(Color online) Sketch for GIFAD from a LiF(001) surface including
relevant angles and channeling directions. Inset: Detail of the (001)
surface depicting the widths $\frac{a}{2}$ and $\protect\delta=\frac{a}{%
\protect\sqrt{2}}$, respectively corresponding to channels $%
\langle100\rangle $ and $\langle110\rangle$.}
\label{Geom1}
\end{figure}

While the feature described in \textit{i)} has already been observed both
experimentally {\cite{Rousseau2008}} and theoretically ~\cite{Muzas2015},
and the one in \textit{ii)} is related to the essentially attractive
character of the VdW interaction, that dominates at long distances from the
surface, the feature given in \textit{iii)} deserves further discussion.
Along the $\langle 110\rangle $ channel, in Ref. ~\cite{Gravielle2008} it
was found that for He/LiF(001) the presence of cationic and anionic rows
(see Fig. \ref{Geom1}) induces polarization within the projectile, resulting
in marked effects on GIFAD patterns. Polarization is implicitly included in
a standard DFT calculation and therefore in our PAW-PBE PES. It mainly
affects the projectile-surface interaction in the medium to large distance
range, where it may compete with VdW interactions, probably overshadowing
them. In contrast, along the $\langle 100\rangle $ channel, neutrally
charged rows of alternating F and Li ions (see Fig. \ref{Geom1}) result in
no projectile polarization effects, and hence in more visible VdW
contributions.

%%%%%%%%%%%%%%%%%%%%%%%%%%%%%%%%%%%%%%%%%%%%%%%%%%%%%%%%%%%%%%%%%%%%%%%%%%%%%%%

\subsection{Simulated GIFAD patterns}

\begin{figure}[tbp]
\includegraphics[width=0.7\textwidth]{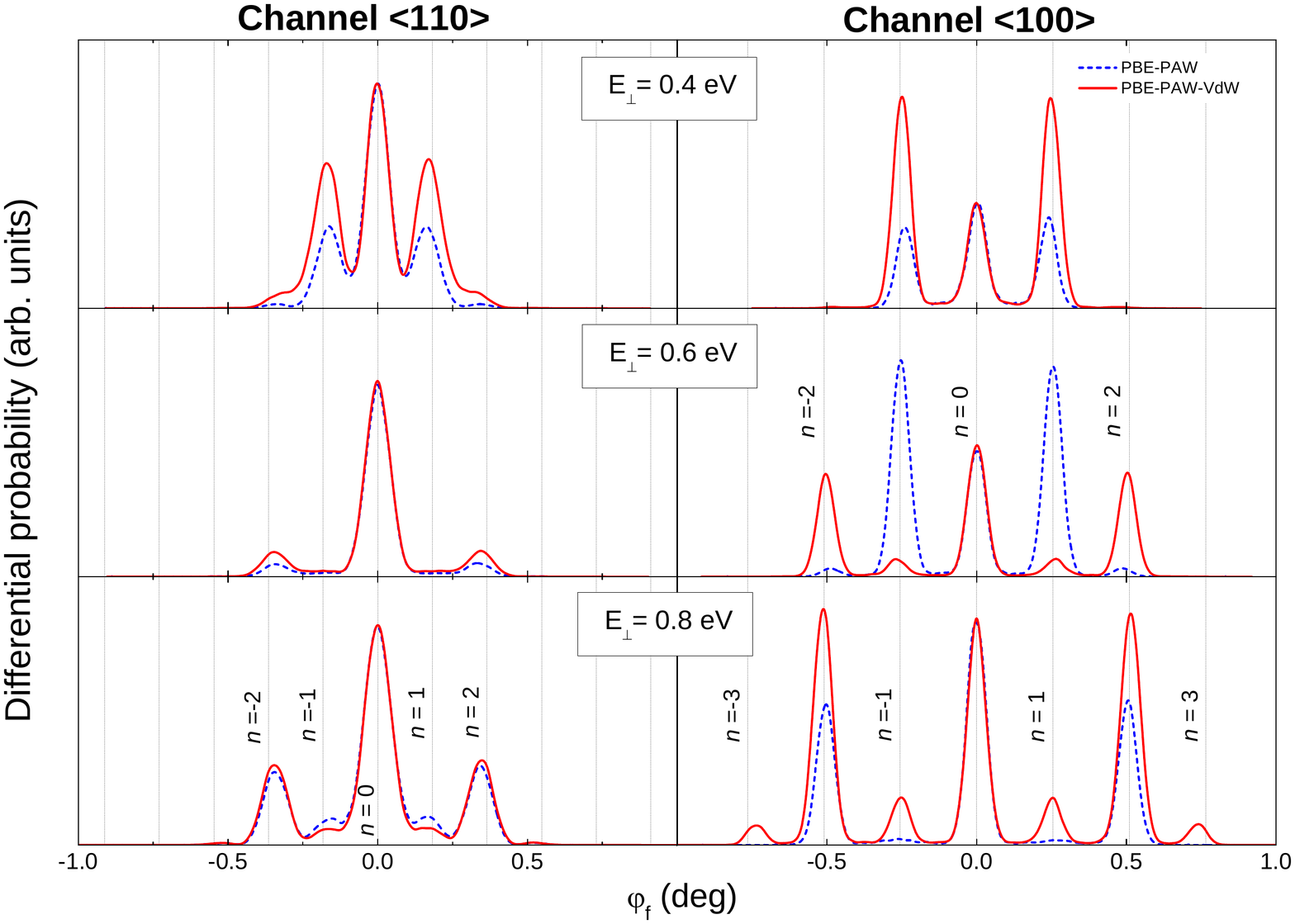}
\caption{(Color online) Simulated GIFAD patterns considering $E_{tot}=1$ keV
and $E_{\perp}$ in the 0.4-0.8 eV range. Left column: Incidence along the $%
\langle110\rangle$ channel. Right column: Incidence along the $%
\langle100\rangle$ channel. Blue dashed line, simulations with the PAW-PBE PES;
red solid line, simulations with the PAW-PBE-VdW PES. The vertical dashed lines indicate the
positions of Bragg peaks, as explained in the text.}
\label{Th_study}
\end{figure}

Simulated GIFAD patterns obtained for both incidence channels and PESs are
displayed in Fig. \ref{Th_study}, as a function of the final azimuthal angle
$\varphi_f$, considering a selection of $E_{\perp}$ values ranging from 0.4
to 0.8 eV.

GIFAD angular distributions present peaks associated with Bragg diffraction,
which are produced by interference among equivalent trajectories whose
starting positions $\overrightarrow{R}_{os}$ lie on different parallel
channels ~\cite{Schuller2008,Winter2011,Gravielle2015b}. These peaks are
situated at azimuthal angles that verify $\sin \varphi _{f}=n\lambda /d$%
, where $\lambda =2\pi /K_{i}$ is the \mbox{de
Broglie} wavelength of the incident atom, $d$ is the width of the
channel ($d=\delta $ for $\langle 110\rangle $ and $d=a/2$ for $%
\langle 100\rangle $), and $n$ is an integer number that determines the
Bragg order. Hence the positions of Bragg peaks, which depend on the total
energy through $\lambda $, provide crystallographic information only. Their
intensities however are determined by a unit-cell form factor that is
originated from interference among trajectories with $\overrightarrow{R}_{os}
$ within the same reduced unit-cell. This unit-cell form factor acts as an
oscillatory envelope function that can reduce or even suppress the
contribution of a given Bragg order ~\cite{Gravielle2015b,Gravielle2014}. In
GIFAD the intensities of the Bragg peaks are extremely sensitive to the
shape of the PES across the incidence channel ~\cite%
{Aigner2008,Gravielle2009,Schuller2010}, being in most cases completely
governed by $E_{\perp }$ ~\cite{Schuller2009,Schuller2009c}. For the present
collision system we have in fact verified that the spectra of Fig. \ref%
{Th_study}, obtained for a fixed incidence energy $E_{tot}=1$ keV, does not
change appreciably upon setting $E_{tot}=3$ keV, while keeping the $E_{\perp
}$ values unchanged. Finally, the number of observed Bragg orders is fully
determined by the unit-cell form factor, which depends on $E_{\perp }$ \cite%
{Muzas2016}.

In Fig. \ref{Th_study} we observe that the PESs features discussed in
subsection III-A directly affect the intensity profiles. Increasing $%
E_{\perp }$ along a given channel results in reflection at more corrugated
regions and thus in wider diffraction patterns while, for fixed $E_{\perp }$%
, the lower corrugation across the $\langle 110\rangle $ channel relative to
that across the $\langle 100\rangle $ one (See Tab. \ref{Tab1}), results in
a narrower pattern for the former. Regarding the VdW contribution, it has a
more visible role along the $\langle 100\rangle $ channel than along the $%
\langle 110\rangle $ one, which is expected due to the difference in the
projectile polarization contribution. Also, for the $\langle 110\rangle $
channel VdW effects grow stronger for lower $E_{\perp }$ values, that is,
when projectiles probe regions farther from the surface. But for\ the $%
\langle 100\rangle $ channel the most marked difference between PAW-PBE and
PAW-PBE-VdW patterns is observed for $E_{\perp }=0.6$ eV. Noteworthy, we
find that, for $E_{\perp }=0.4$ eV, VdW has a marked effect for both
channels.

%%%%%%%%%%%%%%%%%%%%%%%%%%%%%%%%%%%%%%%%%%%%%%%%%%%%%%%%%%%%%%%%%%%%%%%%%%%%%%%

\subsection{Comparison to Experiments}

In this subsection we present simulated GIFAD patterns for different
incidence conditions, corresponding to available experimental data. We
compare to the experiments the results derived from both the PAW-PBE and the
PAW-PBE-VdW PESs, with the aim of assessing their performance in terms of
incidence channel and normal energy. In Tab. \ref{Tab2} we enumerate the
various experimental settings considered in this work, and assign a code
name $(channel\vert E_{\perp}\times100)$ to each of them, with $E_{\perp}$
expressed in eV. We will hereafter use these code names to refer to each
particular setting.

%%%%%%%%%%%%%%%%%%%%%%%%%%%%%%%%%%%%%%%%%%%%%%%%%%%%%%%%%%%%%%%%%%%%%%%%%%%%%%%

\subsubsection{Incidence along the $\langle110\rangle$ channel}

We first present and discuss the simulations and experiments for incidence
along the $\langle110\rangle$ channel, where weaker VdW effects are
predicted.

\begin{figure}[tbp]
\includegraphics[width=0.5\textwidth]{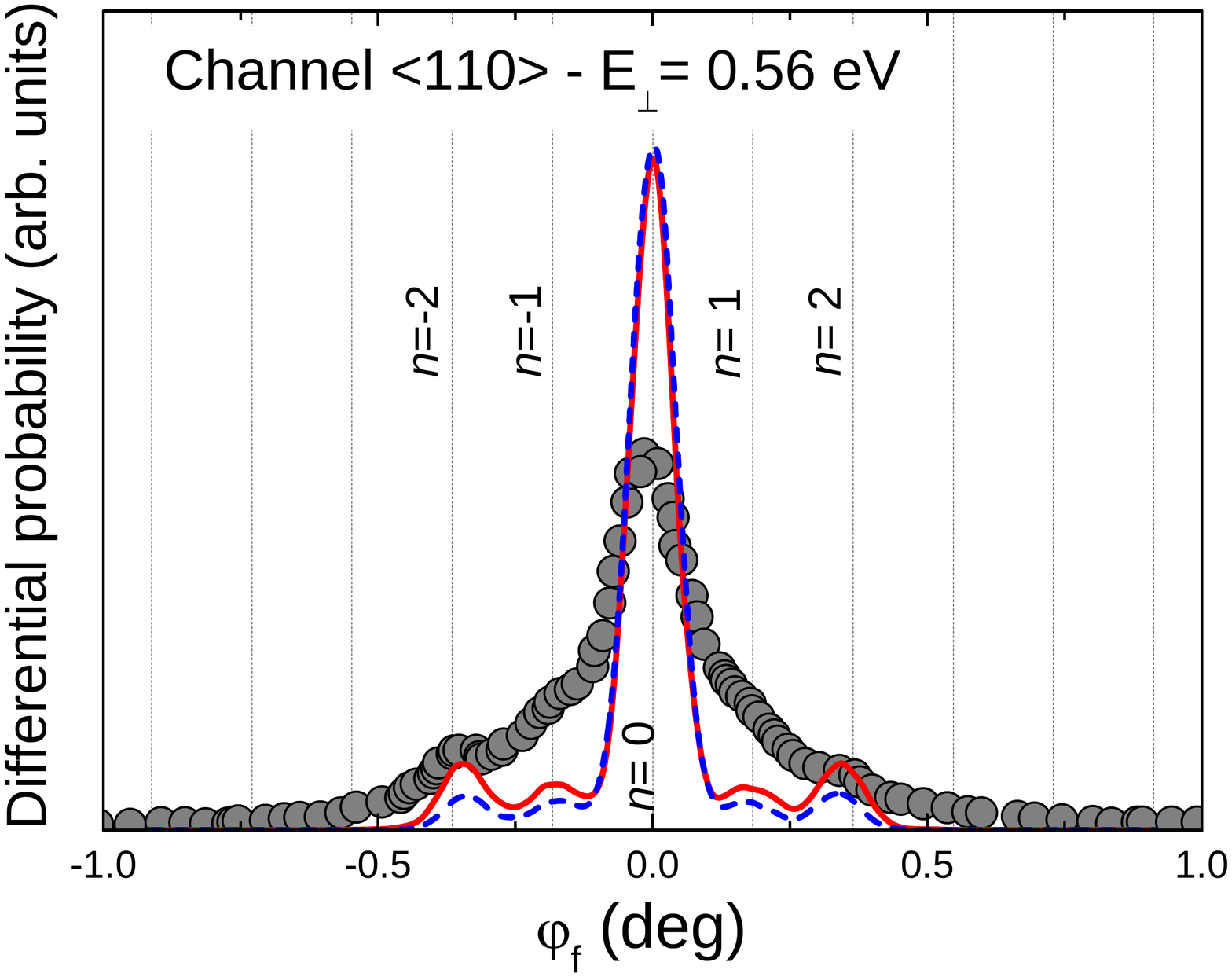}
\caption{(Color online) GIFAD projected intensity profiles for the case $%
(110\vert56)$. Blue dashed line, simulations with the PAW-PBE PES;
red solid line, simulations with the PAW-PBE-VdW PES; solid circles,
experiments from Ref. ~\protect\cite{Rousseau2008}. The vertical
dashed lines indicate the positions of Bragg peaks, as explained in
the text.} \label{A-56}
\end{figure}

In Fig. \ref{A-56}, the simulated intensity distributions for $E_{\perp
}\,=\,0.56$ eV (case $(110|56)$) are contrasted with the experimental
projected intensity profile reported by Rousseau \textit{et al}. ~\cite%
{Rousseau2008}. Theoretical distributions are here normalized to the central
maximum (i.e., at $\varphi _{f}\,=\,0$), while the experimental one is
normalized to the second-order Bragg peak corresponding to the PAW-PBE-VdW
potential. For $(110|56)$, PAW-PBE and PAW-PBE-VdW potentials produce very
similar patterns which satisfactorily reproduce the overall characteristics
of the experimental distribution, showing a very intense central peak sided
by two much lower maxima associated with the $n\,=\,\pm 2$ orders. The small
differences between our PAW-PBE and PAW-PBE-VdW theoretical distributions
can be explained from the values given in Tab. \ref{Tab1}, where the latter
shows a slightly higher corrugation relative to the former, which results in
increased intensities of its nevertheless low $n\,=\,\pm 1,\pm 2$ peaks. At
the positions corresponding to the $n=\pm 1$ peaks, the intensity is higher
in the experiment than in our calculations. However, this discrepancy might
be attributed to experimental limitations that do not allow one to
distinguish the different Bragg orders. At such angular positions the
experimental profile seems to include contributions from the broad central
maximum, which hinder a stringent comparison with the theoretical curves.

\begin{figure}[tbp]
\includegraphics[width=0.5\textwidth]{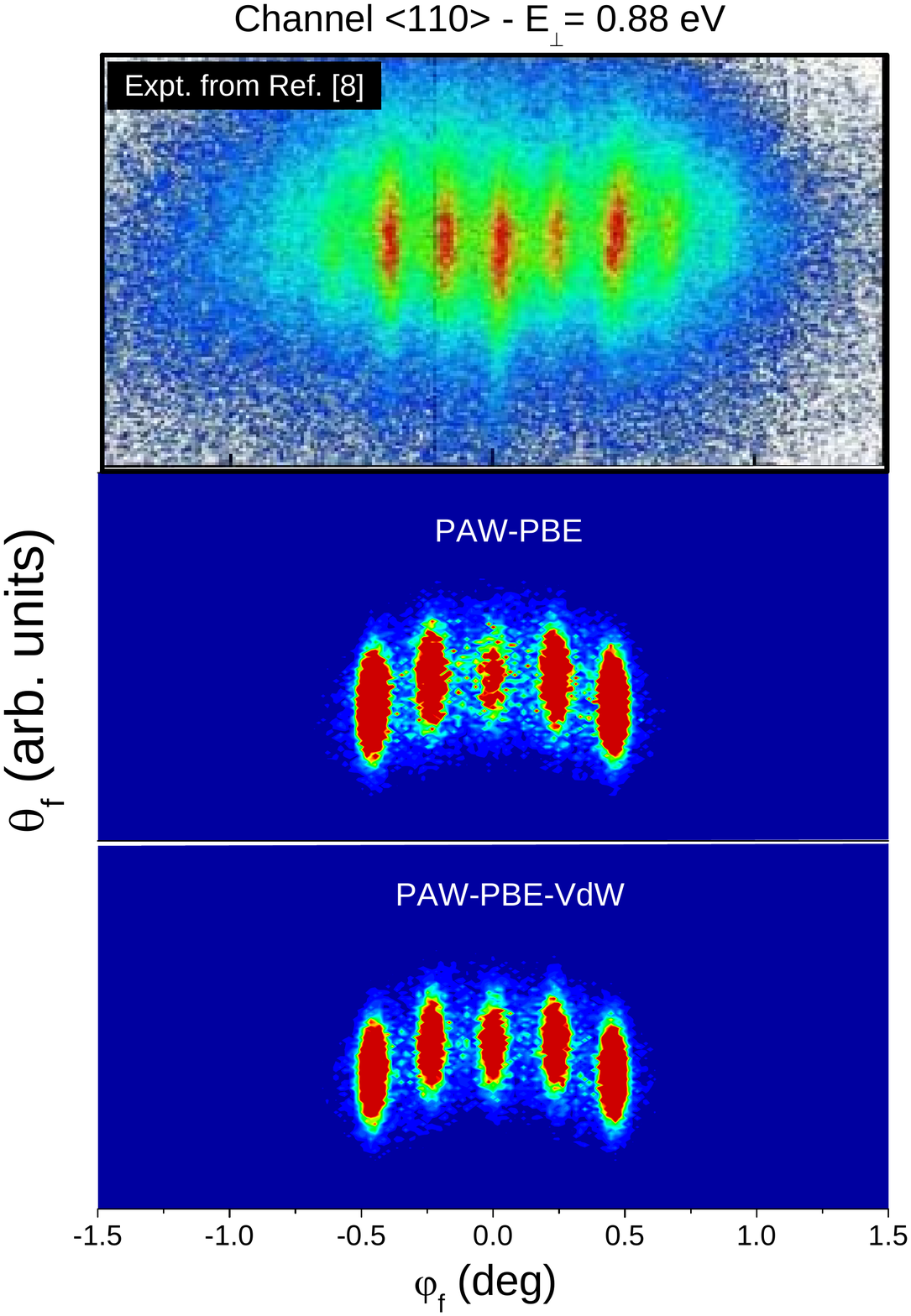}
\caption{(Color online) GIFAD patterns for the case $(110\vert88)$. TOP:
Experiments from Ref. ~\protect\cite{Schuller2007}; MIDDLE: Simulations with
the PAW-PBE PES; BOTTOM: Simulations with the PAW-PBE-VdW PES.}
\label{A-884}
\end{figure}

We continue on to address the higher normal energy $(110|88)$ case,
corresponding to the experiments for \mbox{$E_{\perp}=0.88$ eV}, reported by
Sch\"{u}ller \textit{et al.} ~\cite{Schuller2007}. In Fig.~\ref{A-884} we
show two-dimensional angular distributions, as a function of the final polar
and azimuthal angles $(\theta _{f},\varphi _{f})$. Since the $\theta _{f}$%
-length of GIFAD patterns is affected by the collimating conditions of the
incident beam ~\cite{Gravielle2015,Gravielle2016}, not given in Ref. ~\cite%
{Schuller2007}, the polar angle is here plotted in arbitrary units. Both the
PAW-PBE and PAW-PBE-VdW calculations nicely reproduce the experimental
pattern displaying five maxima with comparable intensities. The similarity
of \mbox{PAW-PBE} and \mbox{PAW-PBE-VdW} GIFAD distributions can be
explained on inspecting the $E_{\perp }\,\sim \,0.88$ eV data in Fig. \ref%
{Equipots}a and Tab. \ref{Tab1}, which show that the corrugations and $Z$%
-ranges for \mbox{PAW-PBE} and \mbox{PAW-PBE-VdW} differ only slightly. The
likeness of both simulated diffraction patterns can thus be traced to that
of the averaged PESs near the reflection region. It is worth noting that,
for this case, the H-Surface reflection distances are only a little larger
than the HF molecule internuclear distance of 0.917 \AA {}.

For incidence along the $\langle110\rangle$ direction, we can then conclude
that our calculations well reproduce the experiments, confirming that
PAW-PBE performs quite satisfactorily while VdW plays a negligible role both
in the PES and in the diffraction patterns for $E_{\perp}\gtrsim0.5$ eV.
Lack of experiments for lower $E_{\perp}$ values prevents us however from
exploring the interesting $E_{\perp}\,\lesssim\,0.4$ eV regime, for which
our theoretical study predicts a visible contribution of VdW interactions.

%%%%%%%%%%%%%%%%%%%%%%%%%%%%%%%%%%%%%%%%%%%%%%%%%%%%%%%%%%%%%%%%%%%%%%%%%%%%%%%

\subsubsection{Incidence along the $\langle100\rangle$ channel}

Following, we address the $\langle 100\rangle $ channel which, according to
our results from Subsection III-A, is more favorable for studying possible
VdW effects, due to the absence of polarization.

\begin{figure}[tbp]
\includegraphics[width=0.5\textwidth]{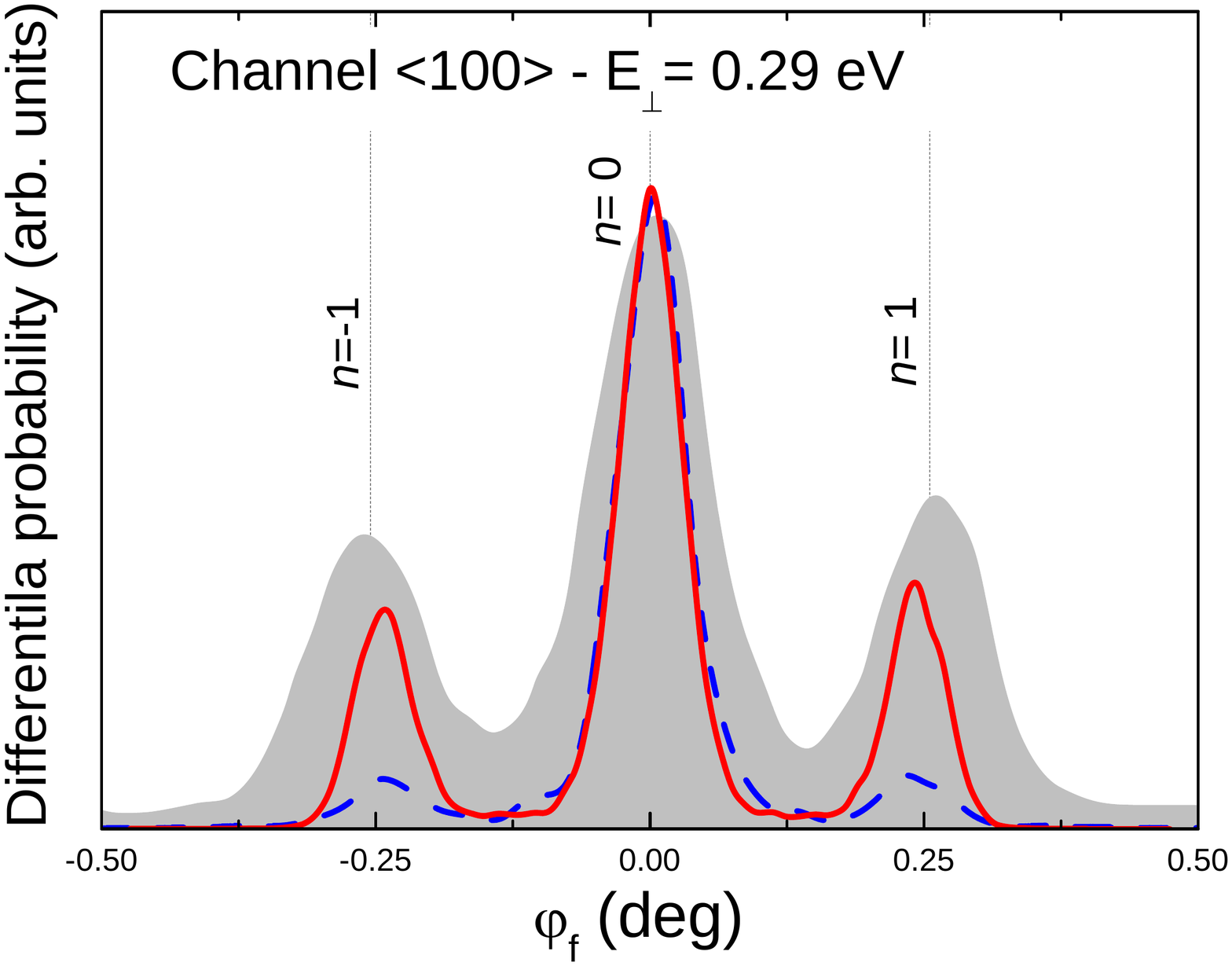}
\caption{(Color online) GIFAD projected intensity profiles for the case $%
(100\vert29)$. Gray shadow line, experiments from Ref. ~\protect\cite%
{Muzas2016}; simulations analogous to Fig. \protect\ref{A-56}. The vertical
dashed lines indicate the positions of Bragg peaks, as explained in the
text. }
\label{B-29}
\end{figure}

In Fig. \ref{B-29} we plot azimuthal angle spectra for the low normal energy
case $(100\vert29)$ ($E_{\perp}\,=\,0.29$ eV). Both our simulated patterns,
as well as the experimental data, taken from Ref. ~\cite{Muzas2016}, are
normalized to the central peak. Remarkably, PAW-PBE results in a rather poor
agreement with the experimental pattern, which strikingly contrasts with the
almost quantitative accord achieved by PAW-PBE-VdW. A key ingredient of this
better performance is the increased corrugation of PAW-PBE-VdW near the
reflection region which, as shown in Fig. \ref{Equipots}b and Tab. \ref%
{Tab1}, approximately doubles that of PAW-PBE.

\begin{figure}[tbp]
\includegraphics[width=0.5\textwidth]{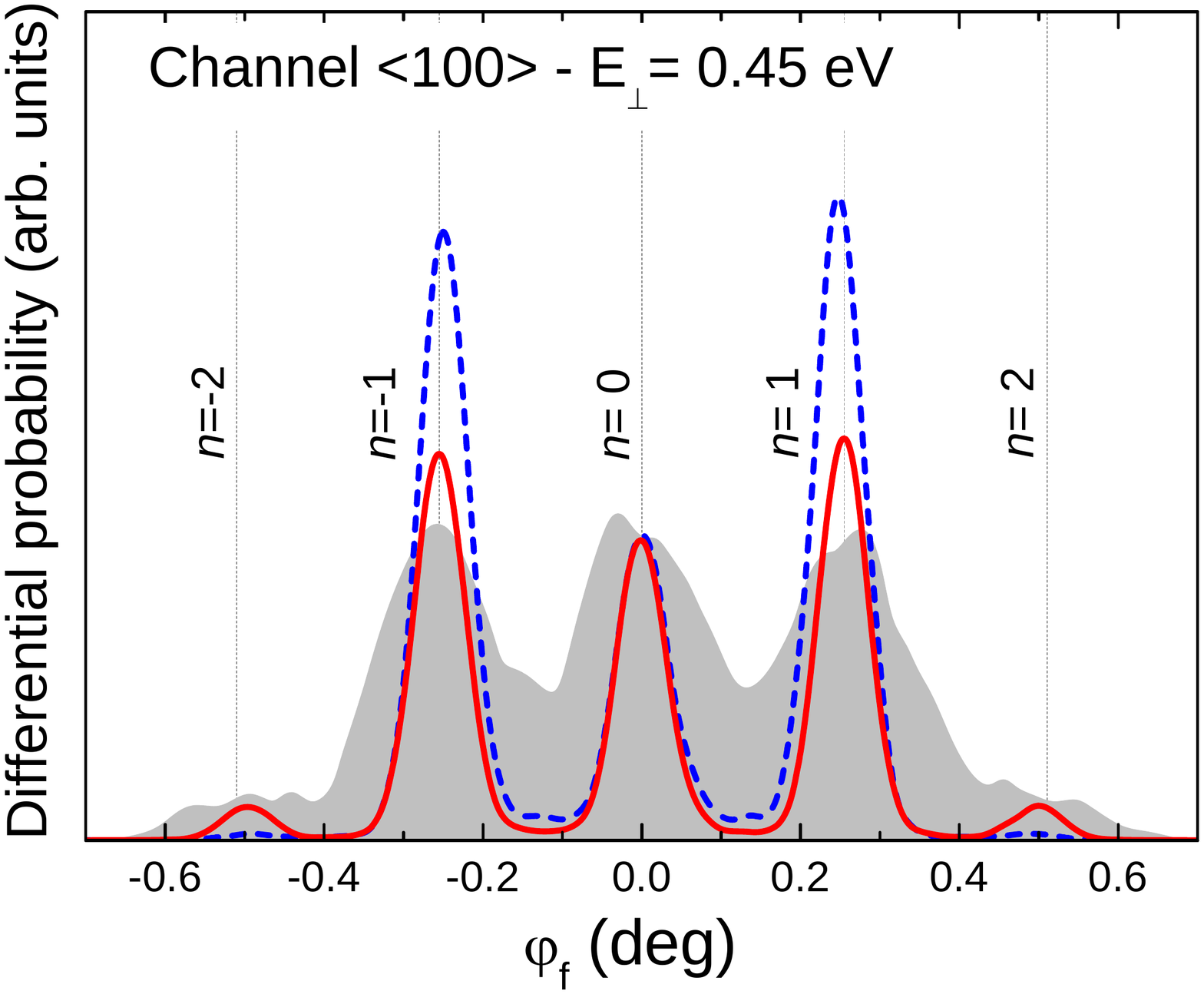}
\caption{(Color online) Analogous to Fig. \protect\ref{B-29} for the case $%
(100\vert45)$.}
\label{B-45}
\end{figure}

\begin{figure}[tbp]
\includegraphics[width=0.5\textwidth]{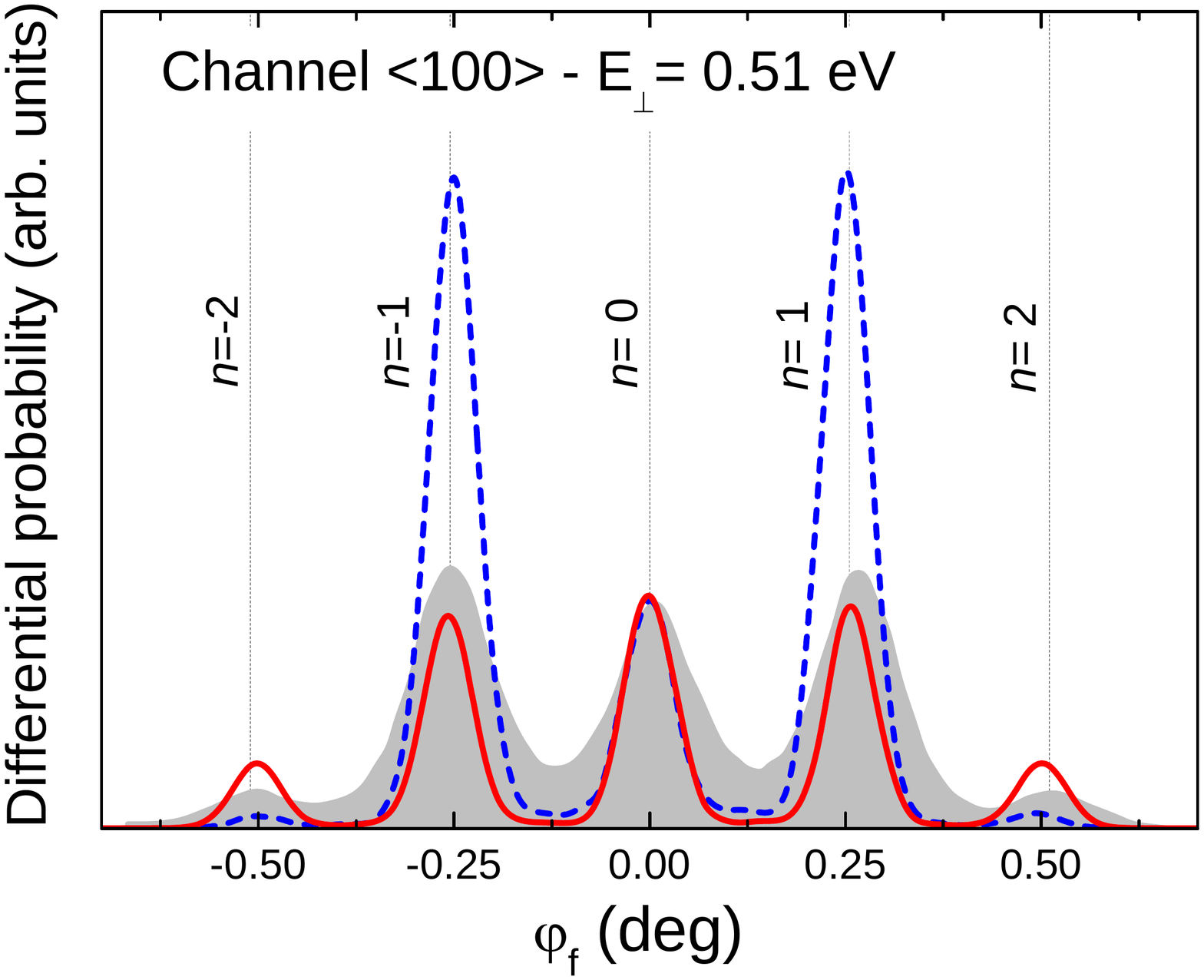}
\caption{(Color online) Analogous to Fig. \protect\ref{B-29} for the case $%
(100\vert51)$.}
\label{B-51}
\end{figure}

In an analogous fashion, the simulated azimuthal angle spectra for cases $%
(100\vert45)$ and $(100\vert51)$ (respectively $E_{\perp}\,=\,0.45$ eV and $%
E_{\perp}\,=\,0.51$ eV), are shown in Figs. \ref{B-45} and \ref{B-51},
together with the corresponding experiments, taken also from Ref. ~\cite%
{Muzas2016}. For both cases, PAW-PBE performs rather unsatisfactorily,
overestimating the $n=\pm1$ Bragg orders, while PAW-PBE-VdW provides a much
better description, reproducing the similar intensities of the $n=0$ and $%
n=\pm1$ orders and the much lower $n=\pm2$ peaks.

Cases $(100|29)$, $(100|45)$ and $(100|51)$ share the common features of
unsatisfactory PAW-PBE patterns and very good descriptions provided by
PAW-PBE-VdW along the complete angular range. In this low-to-intermediate-$%
E_{\perp }$ regime the H atom scattering dynamics is restricted to regions
where the semi-local PBE functional appears to provide an inadequate
description of the H-surface interaction, of which PAW-PBE-VdW seems to give
a much better representation. Concerning a possible influence of the
functional choice on our PES (and hence on the simulated GIFAD patterns) we
can mention that Muzas \textit{et al.} \cite{Muzas2016} have recently
reported GIFAD simulations for the low-to-intermediate $E_{\perp }$ cases
along $\langle 100\rangle $. These authors modeled the projectile-surface
interaction with a PAW-PW91 PES, where the PW91 exchange-correlation
functional was used, instead of the PBE. This PAW-PW91 PES was built with
precision criteria comparable to our PAW-PBE PES and, noteworthy, leads to
GIFAD patterns very similar to our PAW-PBE results. This fact contributes to
support our claim that the reason behind PAW-PBE poor performance for this
channel and normal energy regime is not our functional choice, but the
neglect of VdW interactions.

\begin{figure}[tbp]
\includegraphics[width=0.5\textwidth]{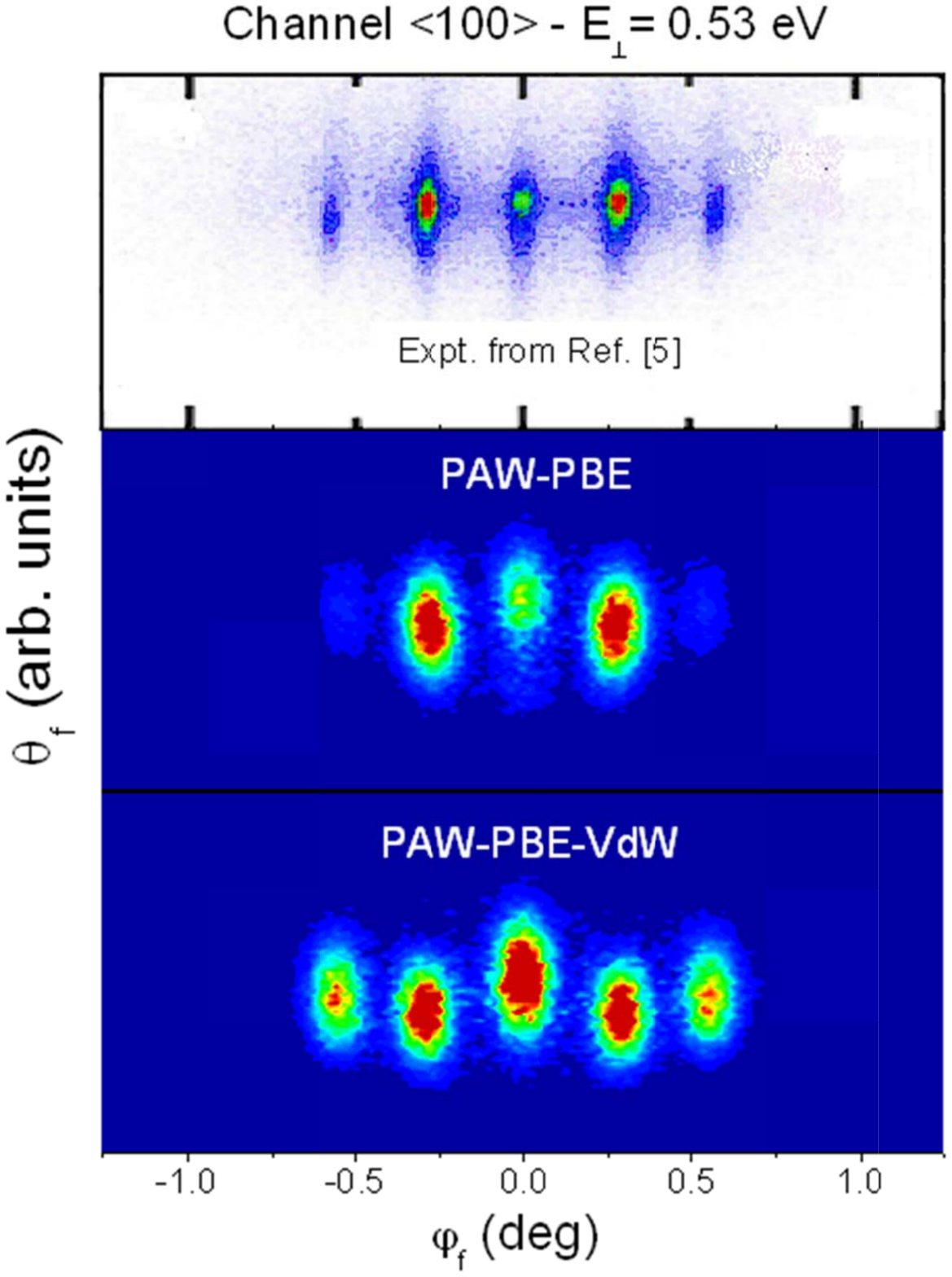}
\caption{(Color online) GIFAD patterns for the case $(100\vert53)$. TOP:
Experiments from Ref. ~\protect\cite{Winter2011}; MIDDLE and BOTTOM panels
analogous to Fig. \protect\ref{A-884}.}
\label{B-53}
\end{figure}

In Fig. \ref{B-53} we show two-dimensional angular distributions in $(\theta
_{f},\varphi _{f})$ for the $(100|53)$ case ($E_{\perp }=0.53$ eV),
corresponding to experiments by Winter \textit{et al.} ~\cite{Winter2011}.
In this case our PAW-PBE calculation performs remarkably well, correctly
reproducing both the outer low-intensity $n=\pm 2$ peaks as well as the
higher intensity of orders $n=\pm 1$ relative to the central $n=0$ peak. On
the contrary, the pattern obtained with PAW-PBE-VdW is not as good, yielding
an apparent overestimation of both the $n=0$ and $n=\pm 2$ peaks.

\begin{figure}[tbp]
\includegraphics[width=0.5\textwidth]{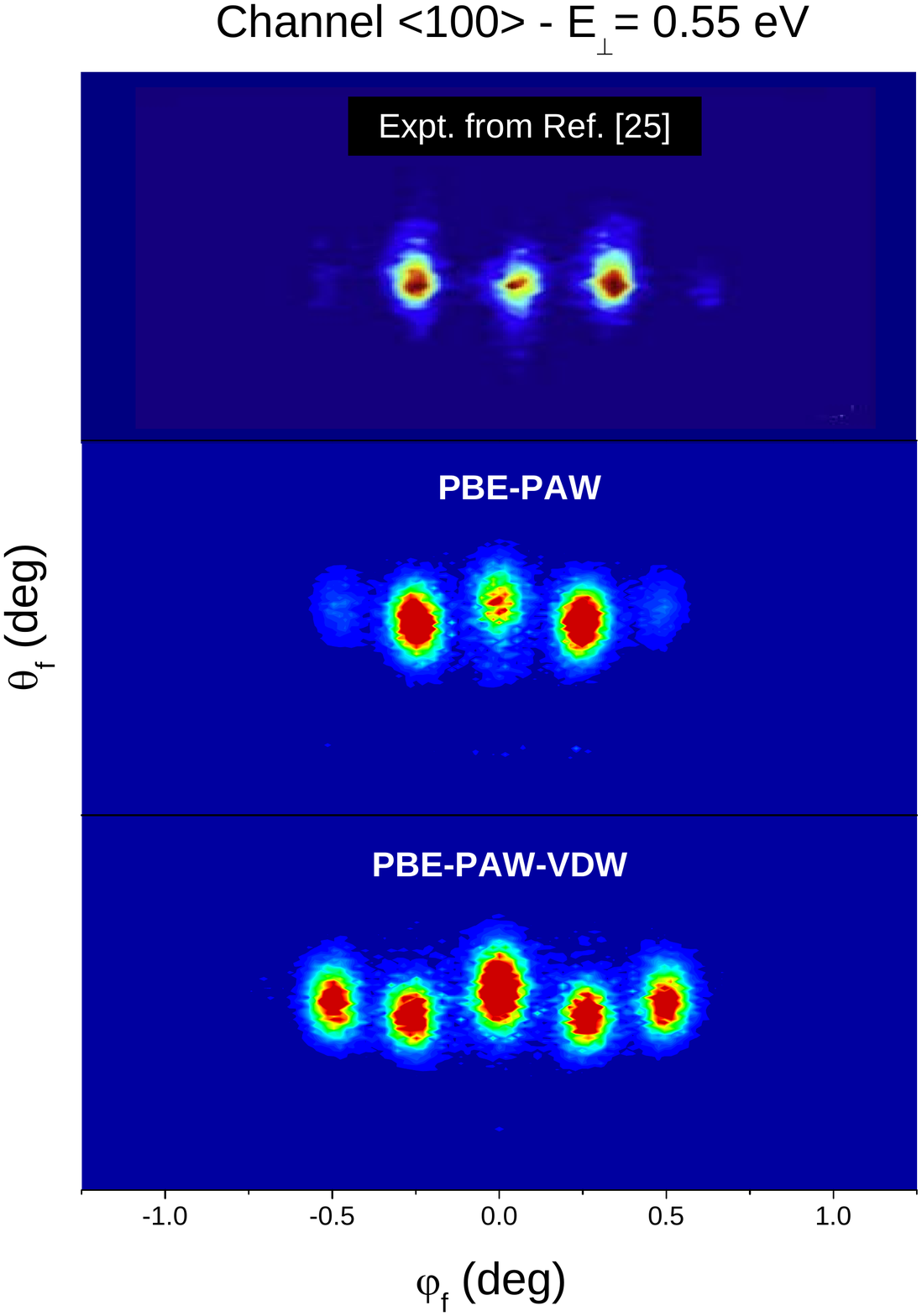}
\caption{(Color online) GIFAD patterns for the case $(100\vert55)$. TOP:
Experiments from Ref. ~\protect\cite{Winter2014}; MIDDLE and BOTTOM panels
analogous to Fig. \protect\ref{A-884}.}
\label{B-55}
\end{figure}

\begin{figure}[tbp]
\includegraphics[width=0.5\textwidth]{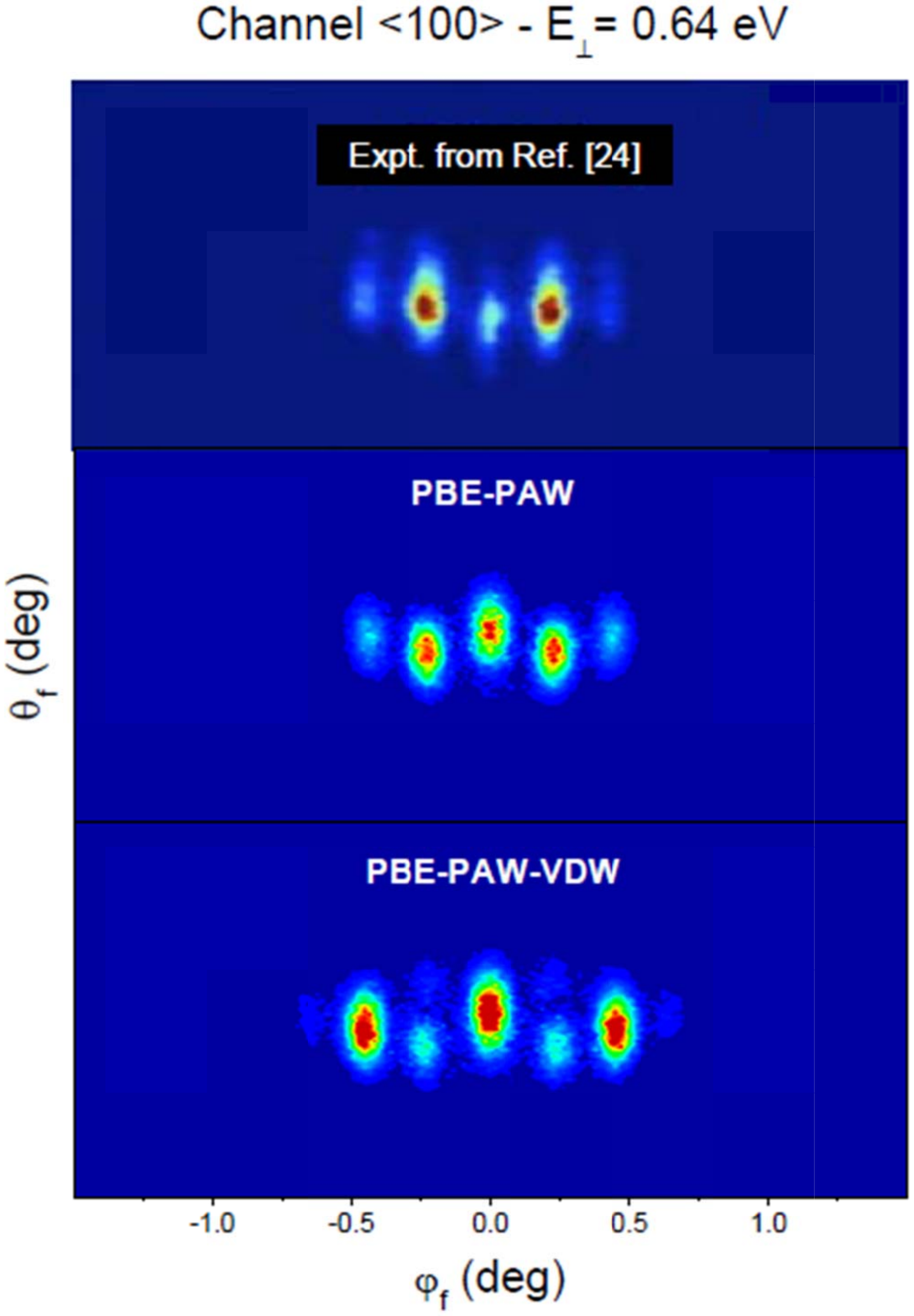}
\caption{(Color online) GIFAD patterns for the case $(100\vert64)$. TOP:
Experiments from Ref. ~\protect\cite{Busch2012}; MIDDLE and BOTTOM panels
analogous to Fig. \protect\ref{A-884}.}
\label{B-64}
\end{figure}

On increasing the $E_{\perp }$ value, the VdW contribution is expected to
gradually fade out. Therefore, it is intriguing that our PAW-PBE-VdW
simulation yields worse-behaved patterns than PAW-PBE ones. This failure is
also observed in Figs. \ref{B-55} and \ref{B-64} for the higher energies $%
E_{\perp }=0.55$ eV and $E_{\perp }=0.64$ eV, respectively, corresponding to
the cases $(100|55)$ and $(100|64)$, and might probably be traced to a
switch-off region issue. In a PES which includes VdW corrections, the
relative importance of semi-local functionals and VdW terms varies with the
H-Surface distance. The matching region, where VdW corrections are smoothly
switched off, is where spurious effects of the VdW approach may arise.
Further increasing the $E_{\perp }$ value results in a convergence of the
PAW-PBE-VdW equipotential curves to the PAW-PBE ones, as shown in Fig. \ref%
{Equipots}, and thus we expect both GIFAD simulations to eventually converge
to a common pattern as well, as was the case for the $\langle 110\rangle ${.}

The very different experimental patterns reported for the $(100|53)$ and $%
(100|51)$ cases illustrate how a small variation in the normal energy may
produce a substantial modification of the GIFAD spectrum. The unexpected
patterns PAW-PBE yields for $E_{\perp }=0.64$ eV, shown in Fig. \ref{B-64},
are probably another example of this. Noteworthy, while PAW-PBE for $0.64$
eV does not satisfactorily reproduce the experimental pattern by Busch
\textit{et al. }\cite{Busch2012}, we do find a nice accord between this
experiment and our PAW-PBE simulation for $0.6$ eV, shown in Fig. \ref%
{Th_study}.

Summing up, for incidence along the $\langle 100\rangle $ channel we
find that, on the one hand, the agreement of PAW-PBE-VdW patterns
with experiments for cases $(100|29)$, $(100|45)$ and $(100|51)$ is
almost quantitative and strongly suggests a non-negligible role of
VdW interactions for low-to-intermediate-$E_{\perp }$ cases in
H/LiF(001) GIFAD. On the other hand, PAW-PBE-VdW performs worse than
PAW-PBE for cases $(100|53)$, $(100|55)$ and $(100|64)$, a feature
that might probably be related to spurious effects of the VdW
approach arising when both the functional and VdW interactions have
non-negligible contributions to the energy. These most interesting
results strongly call for more experiments and theoretical work in
the $E_{\perp }>0.5$ eV energy range.

%%%%%%%%%%%%%%%%%%%%%%%%%%%%%%%%%%%%%%%%%%%%%%%%%%%%%%%%%%%%%%%%%%%%%%%%%%%%%%%

\section{CONCLUSIONS}

In this article, we have studied GIFAD on H/LiF(001), comparing the
diffraction patterns obtained with a PAW-PBE PES with those obtained with a
PAW-PBE-VdW PES, where VdW has been included following Grimme's approach. We
have theoretically investigated the relevance of VdW interactions along the $%
\langle110\rangle$ and $\langle100\rangle$ channels in the 0.4-0.8 eV $%
E_{\perp}$-range and have predicted a marked influence of VdW corrections
for both channels in the low-$E_{\perp}$ ($E_{\perp}\leq 0.4$ eV) regime.
Also we have found that VdW corrections affect more strongly the $%
\langle100\rangle$ than the $\langle110\rangle$ channel, due to the presence
of polarization effects for the latter direction. These effects compete with
VdW in the intermediate to large distance region.

From the comparison with available experiments, we have found that PAW-PBE
gives an adequate description of GIFAD along the $\langle 110\rangle $
channel, for $E_{\perp }\,>\,0.55$ eV cases and, along the $\langle
100\rangle $ channel, for $E_{\perp }\,=\,0.53$ and $0.55$ eV. GIFAD
simulations along the $\langle 100\rangle $ channel for $E_{\perp }\,\leq
\,0.51$ eV cases however result in a poor agreement with experiments unless
VdW interactions are considered, through the PAW-PBE-VdW PES, in which case
our simulated patterns remarkably achieve quantitative agreement with
experiments. This certainly is the main finding of the present work and to
our knowledge, it might be the first evidence of marked VdW effects in
GIFAD, obtained with a DFT potential which includes VdW interactions in a
non ad-hoc fashion, through Grimme's semiempirical approach ~\cite%
{Grimme2006}.

Other worthmentioning points are:\textit{\ i)} The prediction that, for
incidence along the $\langle 110\rangle $ channel, VdW contributions should
become relevant for low impact energies ($E_{\perp }\,\lesssim \,0.4$ eV),
and \textit{ii)} the quite intriguing patterns obtained for the $\langle
100\rangle $ channel in the $E_{\perp }\sim 0.5-0.65$ eV range. More
experiments for both these incidence conditions would be most desirable for
a thorough analysis of the reliability of PES models and the influence of
the VdW interaction.

Moreover, the present results open the way for many further studies on this
topic. In particular, more experiments are important to study the
performance of this semiempirical approach to VdW for $E_{\perp}\,\sim\,0.6$
eV and incidence along the $\langle100\rangle$ channel. The high-$E_{\perp}$
regime is technologically appealing as, under it, GIFAD can reach
topological resolution thus becoming a reciprocal space analog of a perfect
tip AFM ~\cite{Atkinson2015}. Low-$E_{\perp}$ as well as intermediate-$%
E_{\perp}$ GIFAD might prove just as relevant, providing a highly sensitive
quality check for Potential Energy Surfaces, and thus contributing in the
development of an accurate description of VdW interactions within DFT.

%%%%%%%%%%%%%%%%%%%%%%%%%%%%%%%%%%%%%%%%%%%%%%%%%%%%%%%%%%%%%%%%%%%%%%%%%%%%%
%\bibliographystyle{plain}
\bibliographystyle{unsrt}
\bibliography{HLiF_FAD_VdW}

\newpage

\begin{table}[!htp]
\centering
\begin{tabular}{c|c||c|c||c|c}
\multicolumn{2}{c||}{} & \multicolumn{2}{c||}{PAW-PBE} & \multicolumn{2}{c}{
PAW-PBE-VdW} \\ \hline\hline
Channel & E/eV & Z-range/\AA {} & $\Delta Z$/\AA {} & Z-range/\AA {} & $%
\Delta Z$/\AA {} \\ \hline
\multirow{7}{*}{$\langle110\rangle$} & 0.3 & 1.85-1.90 & 0.05 & 1.61-1.71 &
0.10 \\ \cline{2-6}
& 0.4 & 1.60-1.64 & 0.04 & 1.46-1.51 & 0.05 \\ \cline{2-6}
& 0.5 & 1.45-1.49 & 0.04 & 1.35-1.41 & 0.05 \\ \cline{2-6}
& 0.6 & 1.35-1.40 & 0.05 & 1.27-1.33 & 0.06 \\ \cline{2-6}
& 0.7 & 1.27-1.33 & 0.06 & 1.20-1.27 & 0.07 \\ \cline{2-6}
& 0.8 & 1.20-1.27 & 0.07 & 1.15-1.22 & 0.07 \\ \cline{2-6}
& 0.9 & 1.15-1.22 & 0.08 & 1.10-1.18 & 0.08 \\ \hline\hline
Channel & E/eV & Z-range/\AA {} & $\Delta Z$/\AA {} & Z-range/\AA {} & $%
\Delta Z$/\AA {} \\ \hline
\multirow{7}{*}{$\langle100\rangle$} & 0.3 & 1.83-1.90 & 0.07 & 1.56-1.70 &
0.15 \\ \cline{2-6}
& 0.4 & 1.55-1.66 & 0.11 & 1.39-1.55 & 0.16 \\ \cline{2-6}
& 0.5 & 1.40-1.53 & 0.13 & 1.28-1.45 & 0.17 \\ \cline{2-6}
& 0.6 & 1.29-1.43 & 0.15 & 1.20-1.37 & 0.17 \\ \cline{2-6}
& 0.7 & 1.20-1.36 & 0.16 & 1.13-1.31 & 0.18 \\ \cline{2-6}
& 0.8 & 1.13-1.30 & 0.17 & 1.07-1.25 & 0.18 \\ \cline{2-6}
& 0.9 & 1.07-1.25 & 0.18 & 1.02-1.21 & 0.19 \\ \hline
\end{tabular}%
\caption{Approximate reflection range and corrugation for the PESs, channels
and $E_{\perp}$ range considered in this article.}
\label{Tab1}
\end{table}

\begin{table}[tph]
\centering%
\begin{tabular}{c|c|c|c|c}
Channel & $E_{tot}$/keV & $E_{\perp }$/eV & Code Name & Experiments \\
\hline\hline
\multirow{2}{*}{$\langle110\rangle$} & 1 $^{(*)}$ & 0.56 & $(110|56)$ &
Rousseau \textit{et al. } \cite{Rousseau2008} \\ \cline{2-5}
& 0.6 & 0.88 & $(110|88)$ & Sch\"{u}ller \textit{et al. } \cite{Schuller2007}
\\ \hline\hline
\multirow{5}{*}{$\langle100\rangle$} & 1 $^{(*)}$ & 0.29 & $(100|29)$ & \cite%
{Muzas2016} \\ \cline{2-5}
& 1 $^{(*)}$ & 0.45 & $(100|45)$ & \cite{Muzas2016} \\ \cline{2-5}
& 1 $^{(*)}$ & 0.51 & $(100|51)$ & \cite{Muzas2016} \\ \cline{2-5}
& 0.8 & 0.53 & $(100|53)$ & Winter \textit{et al. } \cite{Winter2011} \\
\cline{2-5}
& 1 & 0.55 & $(100|55)$ & Winter \textit{et al. } \cite{Winter2014} \\
\cline{2-5}
& 1.25 & 0.64 & $(100|64)$ & Busch \textit{et al. } \cite{Busch2012} \\
\hline
\end{tabular}%
\caption{Reported experiments for H/LiF(001) GIFAD. Code Name indicates how
we will refer to the particular incidence setting in the text. $^{(*)}$
refers to cases for which the simulations were performed considering $%
E_{tot}=\,1$ keV, verifying the stability of the pattern obtained upon $%
E_{tot}$ variation (keeping $E_{\perp }$ fixed).}
\label{Tab2}
\end{table}

\end{document}